\newcommand{\op}{\mathcal{O}}
\newcommand{\ii}{{\rm i}}
\newcommand{\ex}{{\rm e}}
\newcommand{\bmu} {\bar{\mu}}
\newcommand{\ba}  {\begin{eqnarray*}}
\newcommand{\ea}  {\end{eqnarray*}}
\newcommand{\bc}  {\begin{center}}
\newcommand{\ec}  {\end{center}}
\newcommand{\bi}  {\begin{itemize}}
\newcommand{\ei}  {\end{itemize}}

\newcommand{\Tr}  {\mathop{\rm Tr}}

\def\lsim{\raise0.3ex\hbox{$<$\kern-0.75em\raise-1.1ex\hbox{$\sim$}}}
\def\gsim{\raise0.3ex\hbox{$>$\kern-0.75em\raise-1.1ex\hbox{$\sim$}}}
\documentclass{ws-procs9x6}

\begin{document}

\title{QCD phase diagram at small densities \\ from simulations with 
imaginary $\mu$}

\author{Ph.~de Forcrand}

\address{ETH Z\"urich, 
CH-8093 Z\"urich, Switzerland \\
and CERN, 
CH-1211 Gen\`eve 23, Switzerland \\
E-mail: forcrand@phys.ethz.ch}

\author{O.~Philipsen}

\address{Center for Theoretical Physics, 
Massachussets Institute of Technology, \\
Cambridge, MA 02139, USA \\
E-mail: philipse@lns.mit.edu}  


\maketitle

\abstracts{
We review our results for the QCD phase diagram at baryonic
chemical potential $\mu_B\leq \pi T$.
Our simulations are performed with an imaginary chemical potential $\mu_I$
for which the fermion determinant is positive.
For 2 flavors of staggered quarks,
we map out the phase diagram and
identify the pseudo-critical temperature $T_c(\mu_I)$.
For $\mu_I/T \leq \pi/3$, this is an analytic function,
whose Taylor expansion is found to converge rapidly,
with truncation errors far smaller than statistical ones.
The truncated series may then be continued to real $\mu$,
yielding the corresponding phase diagram for $\mu_B\lsim 500$ MeV.
This approach provides control over systematics and avoids reweighting.
We outline our strategy to find the (2+1)-flavor critical point.
}

\section{Introduction}

Substantial progress has been accomplished over the last two years
toward the numerical determination of QCD properties at finite density.
This progress has not come through a solution of the ``sign problem'',
which prohibits standard Monte Carlo simulations because the Dirac
determinant is no longer real positive at non-zero chemical potential $\mu$.
The standard approach to the sign problem, which requires statistics
growing exponentially with the volume and the chemical potential,
is still current.
Rather, progress has come through a change to a more pragmatic attitude. 
It makes sense to explore the region of {\em small} chemical potential,
with methods whose failure at large $\mu$ has become acceptable.
This is because information so obtained is phenomenologically important, 
especially for heavy-ion collisions where the net quark density remains 
small.

Three approaches are currently pursued, and reviewed in these proceedings:
$(i)$ a two-parameter reweighting, in $T$ and $\mu$, of $\mu=0$ 
simulations~\cite{FK};
$(ii)$ a Taylor expansion of $(i)$ truncated to its lowest non-trivial 
order~\cite{Karsch};
$(iii)$ a study at imaginary $\mu$, followed by analytic continuation
of a truncated Taylor series to real $\mu$.
$(i)$ is limited to small volumes because of the sign problem; $(ii)$
is in addition restricted to smaller chemical potentials, and the systematic
error due to the Taylor truncation is unknown.
We have adopted $(iii)$: it can be used for arbitrarily large
volumes since it has no sign problem, and systematic errors coming
from analytic continuation can be controlled.
We review this approach here, applied to the determination of the
pseudo-critical line $T_c(\mu)$ between confinement and deconfinement~\cite{main}.

\section{QCD at real and imaginary chemical potential}

The grand canonical partition function
$Z(V,\mu,T)=\Tr \left(\ex^{-(\hat{H}-\mu \hat{Q})/T}\right)$
can be considered for a {\em complex} chemical potential
$\mu = \mu_R + \ii \mu_I$. Two general properties suffice to
constrain the phase structure as a function of $\mu_I$: \\
- $Z$ is an even function of $\mu$: $Z(\bmu)=Z(-\bmu)$, 
where $\bar{\mu}=\mu/T$. \\
- A non-periodic gauge transformation, which rotates 
the Polyakov loop by a center element but leaves $Z$ unchanged, 
is equivalent to a shift in $\mu_I$~\cite{Roberge}:
\begin{equation}
Z(\bmu_R,\bmu_I)=Z(\bmu_R,\bmu_I+2\pi/N).
\end{equation}
For QCD ($N=3$), these two properties lead to $Z(3)$ transitions at critical values of
the imaginary chemical potential,
$\bmu_I^c=\frac{2\pi}{3} \left(n+\frac{1}{2}\right)$,
separating regions of parameter space where the Polyakov loop angle 
$\langle\varphi\rangle$ falls
in different $Z(3)$ sectors. Perturbative and strong coupling considerations
led the authors of \cite{Roberge} to predict at $\bmu_I^c$ a first-order
phase transition at high temperature (deconfined phase) and a crossover
at low temperature (confined phase). The resulting phase diagram in the
$(\mu_I,T)$ plane is periodic, and depicted in Fig.~1.

\begin{figure}[htb]
\centerline{\epsfxsize=3.6in\epsfysize=1.8in\epsfbox{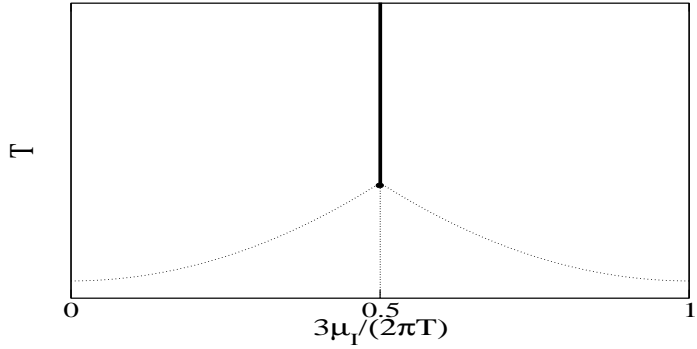}}   
\caption{Schematic phase diagram in the $(\mu_I,T)$ plane. The solid line
marks a first-order $Z_3$ transition.}
\end{figure}

The non-analyticity closest to the origin, at $\bmu_I^c=\frac{\pi}{3}$,
limits the prospects of analytic continuation. 
Our strategy, following \cite{Lombardo-Hart}, consists of fitting by a 
Taylor series in $\bmu_I^2$ observables measured at $(\bmu_R=0,\bmu_I\neq 0)$,
then continuing the truncated Taylor series to real $\bmu$.
Convergence of the Taylor expansion can only be checked for
$\mu_I < \frac{\pi}{3} T$, i.e. $\mu_B \lsim 500$~MeV.

\section{Analyticity of the (pseudo-) critical line}

The phase transition or crossover between confinement and deconfinement,
as $\beta$ (or $T$) is varied while keeping $\mu$ fixed, is characterized
by a peak of the susceptibility 
$\chi= V N_t \left\langle(\op - \langle\op\rangle)^2\right\rangle$ 
for observable $\op$.
The critical line $\beta_c(\mu)$ is thus defined implicitly via
\begin{equation}
\left.\frac{\partial\chi}{\partial\beta}\right|_{\mu,\beta_c}=0,\qquad
\left.\frac{\partial^2\chi}{\partial\beta^2}\right|_{\mu,\beta_c}<0.
\end{equation}
In a finite volume $V$, $\chi(\beta,\mu)$ is analytic, and from the implicit
function theorem $\beta_c(\mu)$ also is.
Moreover, it is an even function of $\mu$ just like $\chi$, so that it
can be expanded in powers of $\mu^2$:
\begin{equation}
\beta_c(\mu)=\beta_c(\mu=0) + \sum_{n=1} c_n (a\mu)^{2n}.
\end{equation}
Analytic continuation between real and imaginary $\mu$ 
(for $|\mu| < \frac{\pi}{3} T$)
is accomplished by flipping the sign of $\mu^2$.

\section{Numerical results for two light flavors}

We have studied QCD with 2 flavors of staggered fermions ($8^3\times 4$ 
lattice, quark mass $a m_q = 0.025$, R-algorithm). The lattice spacing
is $a \sim 0.3$ fm, and sizeable corrections can be expected in the
continuum extrapolation.

The predicted nature of the $Z(3)$ transition at $(a\mu_I)^c=\pi/12$
is confirmed,
and the phase diagram is that of Fig.~1. The confinement-deconfinement
transition line is obtained from the peaks of the plaquette, quark condensate
and Polyakov loop susceptibilities, which give consistent values $\beta_c(a\mu_I)$
(see Fig.~2). These values are then fitted by a Taylor series in $(a\mu_I)^2$
over the interval $a\mu_I \in [0,\pi/12]$. The data are well described by
a linear fit, and agreement hardly improves with a quadratic fit (see Fig.~3a).
The quadratic coefficient $\propto (a\mu_I)^4$ is zero within errors.
Therefore, we can safely approximate the critical line as
\begin{equation}
\beta_c(a\mu_I) = 5.2865(18) + 0.596(40) (a\mu_I)^2 \quad .
\end{equation}
We translate into physical units using the perturbative
two-loop $\beta$-function, which suffices for our present accuracy. 
This yields, as a function of the baryonic chemical potential $\mu_B$
\begin{equation}
\frac{T_c(\mu_B)}{T_c(\mu_B=0)}= 1 - 0.00563(38) \left(\frac{\mu_B}{T}\right)^2,
\end{equation}
using $T_c(\mu=0)=173(8)$ MeV,
while the next-order term $\op((\mu_B/T)^4)$ is statistically insignificant
up to $\mu_B \sim 500$ MeV (see Fig.~3b).
Similar results have been obtained for 4 flavors~\cite{DElia}.

\section{How to find the (2+1)-flavor critical endpoint}

In the case of 3 flavors (degenerate or not), the deconfinement ``transition''
changes from crossover to first-order at a critical point $(T_c,\mu_c)$,
which can be identified by a study of cumulant ratios or Lee-Yang zeroes.
As the quark mass $m_q$ is varied, the critical point describes an
analytic curve $T_c(\mu)$ (Fig.~4), which can be Taylor expanded. 
On the lattice:
\begin{equation}
\beta_c(\mu)=\beta_c(\mu=0) + \sum_{n=1} b_n (a\mu)^{2n} \quad .
\end{equation}
As in Section 4, data can be obtained at imaginary $\mu$, then analytically
continued to real $\mu$ by flipping the sign of $\mu^2$.

\begin{figure}[t]
\centerline{\epsfxsize=2.3in\hspace*{0cm}\epsfbox{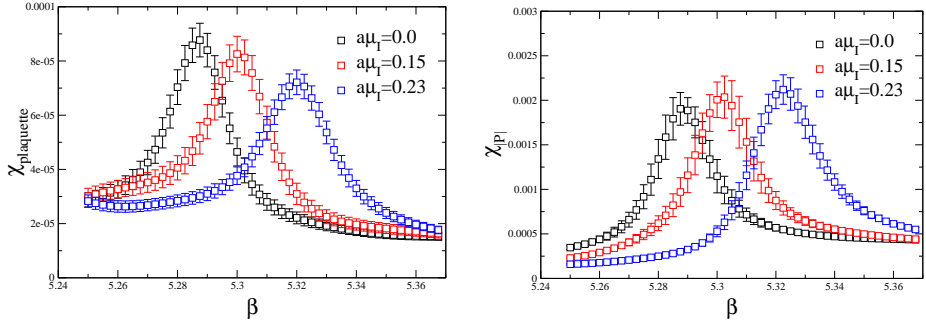}%
            \epsfxsize=2.3in\hspace*{0.5cm}\epsfbox{absPsus.eps}}
\caption{Susceptibilities of the plaquette ({\em left}) and of the magnitude of the
Polyakov loop ({\em right}) for various $\mu_I$'s, as a function of $\beta$.}
\end{figure}

\begin{figure}[h]
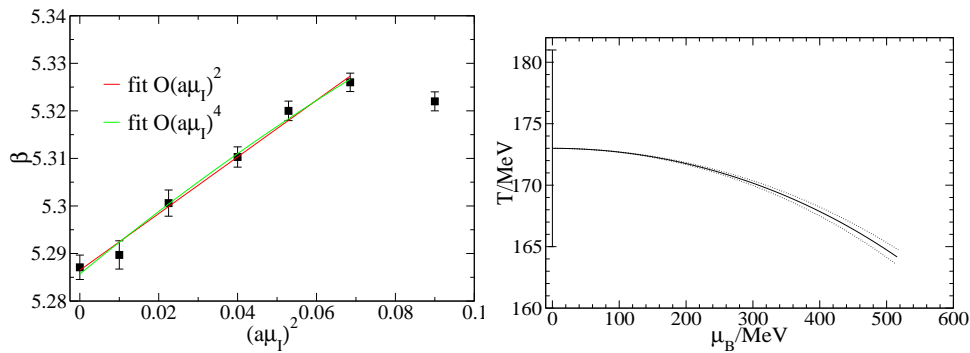

\centerline{\epsfxsize=2.5in\epsfysize=1.8in\epsfbox[0 16 723 532]{pdiag.eps}
            \epsfxsize=2.48in\epsfysize=1.65in\epsfbox[17 23 728 522]{rpdiag.eps}}
\caption{({\em left}) Phase diagram in the $((a\mu_I)^2,\beta)$ plane. The rightmost point
lies beyond the $Z_3$ transition, and is therefore excluded from the fit.
({\em right}) Analytically continued phase diagram in the $(\mu_B,T)$ plane.}
\end{figure}

\begin{figure}[h]
\leavevmode
\centerline{\epsfxsize=3.0in\epsfysize=2.0in\vspace*{0cm}\epsfbox{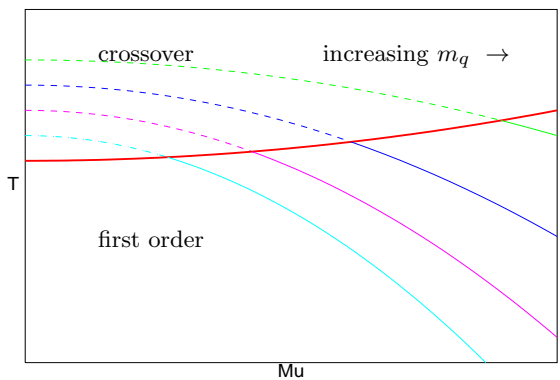}}
\put(-235,50.0){\small first order}
\put(-235,120){\small crossover }
\put(-150,120){\small increasing $m_q~\rightarrow$}
\caption{Critical lines in the $(T,\mu)$ plane for different quark masses $m_q$. The bold upward
parabolic curve characterizes second-order transition points, separating 
the crossover and the first-order regimes. It can
be analytically continued to negative $\mu^2$.}
\end{figure}

\end{document}